\documentclass[preprint,showpacs,showkeys,floatfix,preprintnumbers,amsmath,amssymb]{revtex4}

\usepackage{graphicx}
\usepackage{dcolumn}
\usepackage{bm}
\usepackage{verbatim}
\usepackage[english]{babel}
\usepackage{color}

\newcommand{\LCMO}{La$_{0.7}$Ca$_{0.3}$MnO$_{3}$}
\newcommand{\LCMOx}{La$_{1-x}$Ca$_{x}$MnO$_{3}$}
\newcommand{\TMI}{$\rm{T_{MI}}$}

\begin{document}

\title{Fingerprint of dynamical charge/spin correlations in the tunneling spectra of colossal
    magnetoresistive manganites}

\author{S. Seiro}
\altaffiliation[Present address: ]{ Max Planck Institute for
Chemical Physics of Solids, N\"{o}thnitzer Str. 40, 01187 Dresden,
Germany. E-mail: seiro@cpfs.mpg.de}
\author{Y. Fasano}
\altaffiliation[Present address: ]{Instituto Balseiro and Centro
At\'{o}mico Bariloche, Bustillo 9500, 8400 Bariloche, Argentina.}
\author{I. Maggio-Aprile}
\author{E. Koller}
\author{R. Lortz}
\altaffiliation[Present address: ]{Department of Physics, The Hong
Kong University of Science \& Technology, Hong Kong. }
\author{{\O}. Fischer}

\affiliation{%
D\'epartement de Physique de la Mati\`ere Condens\' ee,
Universit\'e de Gen\`eve, \\Quai Ernest-Ansermet 24, 1211 Geneva,
Switzerland
}%


\begin{abstract}

We present temperature-dependent scanning tunneling spectroscopy
measurements on \LCMOx\ ($x\sim0.33$) films with different degrees
of biaxial strain. A depletion in normalized conductance around
the Fermi level is observed both above and below the
insulator-to-metal transition temperature \TMI, for weakly as well
as highly-strained films. This pseudogap-like depletion globally
narrows on cooling. The zero-bias conductance decreases on cooling
in the insulating phase, reaches a minimum close to \TMI\ and
increases on cooling in the metallic phase, following the trend of
macroscopic conductivity. These results support a recently
proposed scenario in which dynamical short-range
antiferromagnetic/charge order correlations play a preeminent role
in the transport properties of colossal magnetoresistive
manganites [R. Yu \textit{et al}., Phys. Rev. B \textbf{77},
214434 (2008)].

\end{abstract}

\pacs{71.30.+h,75.47.Lx,68.37.Ef,73.50.-h}

\keywords{metal-insulator transition, manganites, scanning tunneling
microscopy, thin films}
\maketitle

In spite of extensive theoretical and experimental research on
colossal magnetoresistive manganites, the mechanism underlying the
transition from insulator to metal-like transport concomitant to
ferromagnetic ordering has not yet been completely
understood~\cite{intro}. In the paramagnetic (PM) phase, tunneling
spectroscopic measurements are in agreement with the presence of a
gap in the density of states: Tunneling conductance presents a
depletion at low bias voltages~\cite{Seiro08,Singh08} and the
zero-bias conductance follows a thermally-activated-like
behavior~\cite{Ronnow06,Singh08}. On cooling into the ferromagnetic
state the depletion does not disappear~\cite{Seiro08,Singh08}, in
apparent contrast to the macroscopic transport properties. In this
work we show that for \LCMO\ (LCMO) films of different strain levels
the depletion observed in both the insulating and metallic regimes
is accompanied by a spectral weight redistribution on cooling
through \TMI. The temperature evolution of the zero-bias conductance
(ZBC) accounts for the macroscopic insulator-to-metal transition.
Our results support a recent theoretical study which shows that
dynamical nanoscale antiferromagnetic/charge order (AFM/CO)
correlations give rise to a pseudogap in the density of states (DOS)
around the chemical potential ($\mu$), not only above but also below
the insulator-to-metal transition temperature \TMI~\cite{Yu08}.

The films studied in this work have been grown by rf sputtering on
(100) SrTiO$_3$ (STO) and (110) NdGaO$_3$ (NGO). The growth
procedure has been reported in detail in
Refs.~\cite{Seiro07APL,Seiro08}. X-ray diffraction measurements
confirmed structural homogeneity, single crystallinity, and the
presence of a single phase. Both LCMO/STO and LCMO/NGO films were
found to be under strain. Reciprocal space mapping for LCMO/STO
films confirmed an in-plane parameter equal to that of the substrate
and a strongly reduced out-of-plane lattice parameter ($c\sim
3.80$\,\AA)~\cite{Seiro08}. The LCMO/NGO film is weakly strained
($c\sim 3.87$\,\AA) and compressed on average in the
plane~\cite{Seiro07APL}. As estimated from Laue oscillations, the
thickness of the films is well above the dead-layer
thickness~\cite{Bibes02,Mueller02}.

The resistivity ($\rho$) of the films was measured in a four-point
configuration, with the current flowing parallel to the film plane,
along one of the main pseudocubic axis. Both kind of films exhibit a
transition from insulator- to metal-like behavior at a temperature
$\rm{T_{MI}}$ lower than that of bulk compounds, as expected for
films under biaxial
strain~\cite{Seiro07APL,Vengalis00,Song02,Bibes02,Ziese03,Valencia03}.
For LCMO/STO, $\rm{T_{MI}}$=154\,K while for LCMO/NGO,
$\rm{T_{MI}}$=235\,K. The residual resistivities are
5\,m$\Omega\,$cm and 0.35\,m$\Omega\,$cm, respectively. By taking
advantage of the shift in $\rm{T_{MI}}$ due to substrate-induced
strain~\cite{Seiro07APL,Vengalis00,Song02,Bibes02,Ziese03,Valencia03},
we accessed the metallic and insulating phases in a wide range of
temperatures.

Since three-dimensional perovskites lack an easy cleaving plane, we
thoroughly cleaned the surface of the films with isopropanol in an
ultrasonic bath prior to entering the sample in a
variable-temperature home-made scanning tunneling microscope. This
procedure has been widely applied to non-cleavable manganites
~\cite{Seiro08, Sudheendra07, Rossler08}. Topographs and current vs.
voltage, $I(V)$, maps were measured as a function of temperature.
Topographs were acquired at a constant current of 0.2-0.6\,nA and a
bias voltage of 1-3\,V. Spectroscopy was measured at a fixed
tip-sample separation (0.5\,nA at 0.5\,V bias) by recording the
current when ramping the bias voltage. The tip electrode made of
electrochemically etched Ir was grounded and the bias voltage $V$
was applied to the sample.

\begin{figure*}[ht!]
\includegraphics[width=1.0\textwidth]{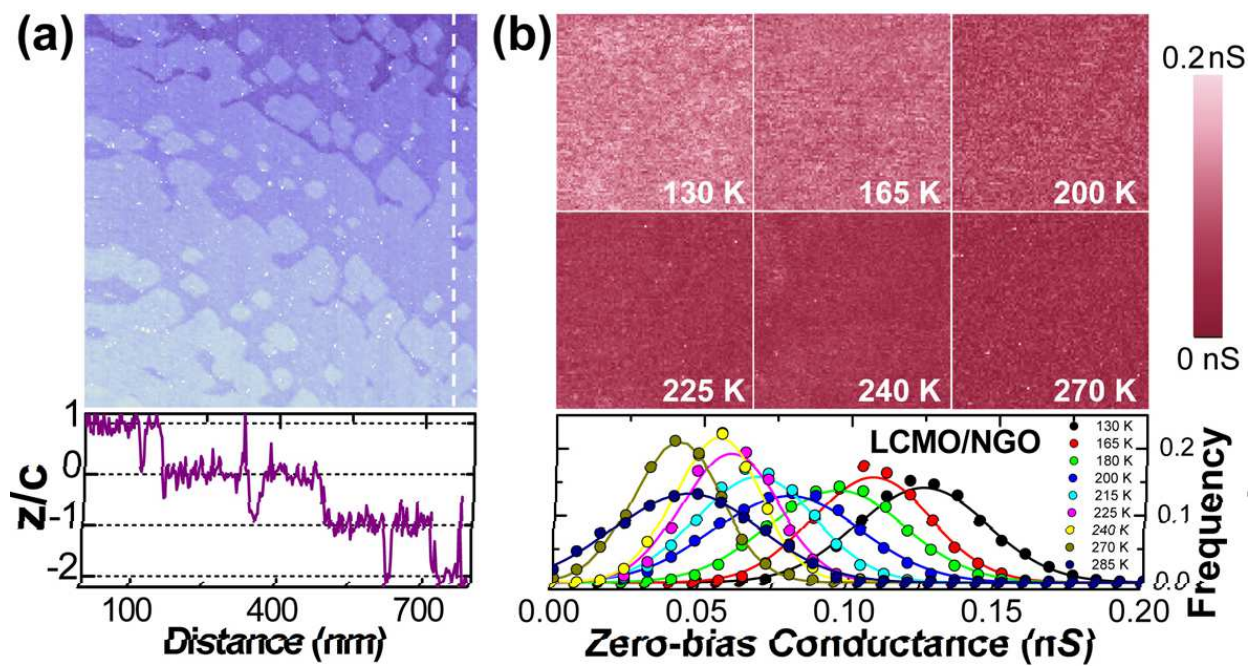}
\caption{\label{Figure1} Topographic and spectroscopic properties of
weakly-strained LCMO/NGO. (a) Top: 800$\times$800\,nm$^2$ topograph
taken at 258\,K for a bias voltage of 1.5\,V and a tunnel current of
0.2\,nA. Bottom: Topographic profile along the dashed line. (b) Top:
450$\times$450\,nm$^2$ zero-bias conductance maps taken at different
temperatures for a junction impedance of 1\,G$\Omega$ (0.5\,V,
0.5\,nA). The color scale used is the same for all maps. Bottom:
Distribution of zero-bias conductance values. The full lines are
Gaussian fits to the experimental distribution.}
\end{figure*}

Topographs reveal flat terraces separated by growth steps multiple
of the out-of-plane pseudocubic lattice parameter for both LCMO/NGO
and LCMO/STO films~\cite{Seiro08,Seiro07JMMM}. In
Fig.~\ref{Figure1}(a) we show a typical topograph for the 42\,nm
thick LCMO/NGO film. ZBC maps over an area covering a few
topographic steps exhibit some dispersion of conductance values but
no particular spatial pattern, as shown in Fig.~\ref{Figure1}(b) for
LCMO/NGO and in Refs.~\cite{Seiro07JMMM,Seiro08Thesis} for LCMO/STO.
For both LCMO/NGO and LCMO/STO films the ZBC distribution is
Gaussian at all temperatures, complementing our previous results at
higher energies on LCMO/STO films~\cite{Seiro07JMMM,Seiro08}. Note
that the ZBC level of a pixel in Fig~\ref{Figure1}(b) is not the
average of ZBC values over the pixel suface, but the point value at
the pixel center. If there were sub-pixel regions of very high and
very low ZBC, as observed for Pr-doped LCMO films at 1.5-2\,eV
energies ~\cite{Ma05}, they should be manifest in the distribution
of conductance values with a statistical weight proportional to
their area fraction. The absence of a bimodal distribution of
conductance values rather appears to indicate that there is no
\textit{static} phase separation in regions with strongly
differentiated electronic properties. Assuming that a ``different''
phase has been missed by the discrete positioning of the tip, the
surface fraction of it would be limited to roughly $10^{-4}$,
casting a doubt on the role it plays in transport properties. The
absence of regions with strongly differentiated tunneling
spectroscopic properties (i.e. of a bimodal ZBC distribution) on
LCMO films has also been reported for LCMO/NGO at
\TMI~\cite{Mitra05} and for cation ordered  LCMO/MgO ($x=1/4$)
 at 115\,K and 294\,K~\cite{Sudheendra07}.

\begin{figure}[bbb]
\includegraphics[width=0.5\textwidth]{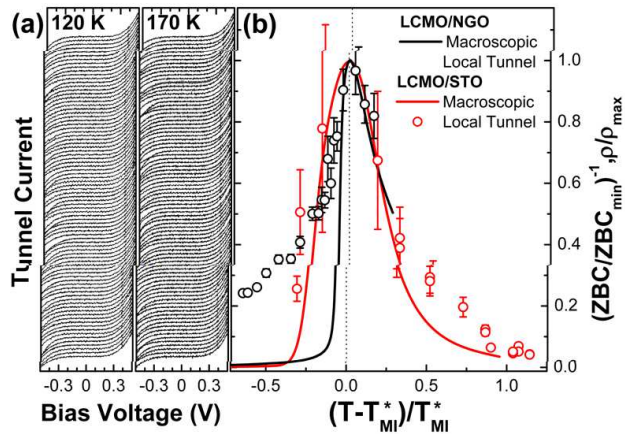}
\caption{\label{Figure2} (a) Local $I(V)$ curves acquired along a
350\,nm trace on highly-strained LCMO/STO for the metallic (120\,K)
and insulating (170\,K) phases. (b) Local (open symbols) vs.
macroscopic (lines) electronic properties vs. temperature for
LCMO/NGO (black) and LCMO/STO (red). Temperatures are expressed
relative to the respective insulator-to-metal transition
temperatures: $\rm{T_{MI}^{*}=T_{MI}}$ for resistivity curves, while
for zero-bias conductance $\rm{T_{MI}^{*}=T_{MI}^S}$, the
temperature at which ZBC reaches a minimum. Since scanning tunneling
spectroscopy probes the DOS at the surface, $T_{MI}^S$ is slightly
different to $T_{MI}$.}
\end{figure}

For both strain states, not only the ZBC values but also the whole
$I(V)$ curves at any given temperature display no bimodal
distribution over the field of view. As an example,
Fig.~\ref{Figure2}(a) shows local LCMO/STO $I(V)$ curves acquired
over a 350\,nm-long path at temperatures below and above the
transition. The current vs. voltage curves in the metallic and
insulating phases present a highly non-linear character and are
remarkably alike. At first sight, one might be tempted to link these
observations to insulating behavior (DOS$(\mu)=0$). However,
non-linear $I(V)$ characteristics can appear also in metals
(DOS$(\mu)\neq0$), in particular if the DOS is non-constant in the
vicinity of $\mu$, or if the tunneling barrier is energy-dependent
(e.g. for bias voltages that are a non-negligible fraction of the
apparent workfunction~\cite{barrier}). Further information on the
DOS$(\mu)$ can be gained by a close inspection of the behavior of
zero-bias conductance as a function of temperature.

Figure~\ref{Figure2}(b) presents the temperature evolution of
macroscopic resistivity and the inverse of the ZBC obtained from the
spatially-averaged local $I(V)$ curves, for both weakly and
highly-strained films. We found that 1/ZBC is finite at all measured
temperatures and roughly follows the behavior of macroscopic
resistivity, increasing on warming in the low-temperature phase and
decreasing on warming in the high-temperature phase. For a gapped
DOS, a thermally-activated ZBC is expected, the activation energy
yielding an estimate of the gap. In the high temperature phase, the
activation energy obtained from the slope of our
$\log(\mathrm{ZBC})$ data vs. $1/\mathrm{T}$ is comparable to the
activation energy estimated from the slope of $\log(\rho)$ vs.
$1/\mathrm{T}$, $(0.12\pm0.01)$\,eV for LCMO/STO and
$(0.07\pm0.01)$\,eV for LCMO/NGO. However, the thermal activation
here may be only apparent. The computational study in
Ref.~\cite{Yu08} proposes the occurrence of \textit{dynamic}
nanoscale AFM/CO correlations in the vicinity of the FM-AFM/CO phase
boundary. If the characteristic lifetime of these correlations is
shorter than the timescale of tunneling experiments, they will not
be visualized in ZBC images. Nevertheless the presence of these
correlations yields a finite DOS ($\mu$) at all temperatures for
long Monte Carlo time scales in the calculation~\cite{Yu08}, with a
temperature dependence that follows the trend of macroscopic
conductivity. This is indeed observed in our films, suggesting that
the observed activated-like behavior does not come from a thermal
excitation of carriers over a band gap but is due to dynamical
CO/AFM correlations.

\begin{figure}[bbb]
\includegraphics[width=0.5\textwidth]{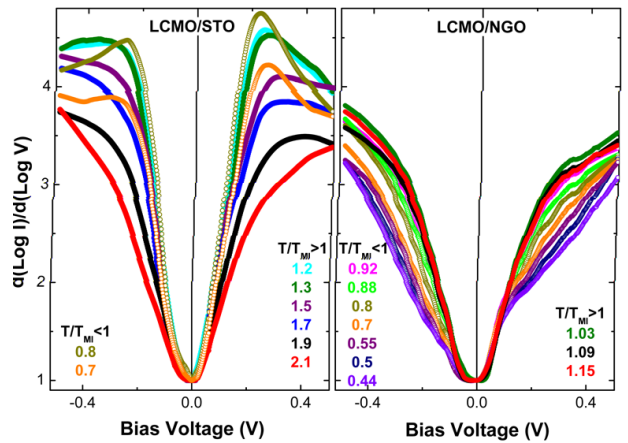}
\caption{\label{Figure3} Normalized conductance curves at different
temperatures both above (full symbols) and below (open symbols) the
insulator-to-metal transition for LCMO/STO (left panel) and LCMO/NGO
(right panel).}
\end{figure}

The DOS in Ref.~\cite{Yu08} is not gapped above \TMI, but presents a
pseudogap-like depletion in the vicinity of $\mu$ with a pseudogap
energy scale that globally decreases on cooling. We can directly
test these predictions by calculating the normalized or logarithmic
conductance, $(dI/dV)/(I/V)$, a method that attenuates the
dependence on the tunneling barrier and yields a quantity
proportional to the local DOS~\cite{Feenstra94}. Normalized
conductance (NC) curves are presented in Fig.~\ref{Figure3} for both
highly (left panel) and weakly (right panel) strained films. At high
temperatures NC curves present a depletion around zero bias. This
depletion survives down to low temperatures, but NC peaks (in the
case of LCMO/STO) or kinks (for LCMO/NGO) become increasingly marked
upon cooling. These features have been previously ascribed to the
spectral signature of polarons~\cite{Seiro08, Wei97}, the distance
between the peaks being related to the polaron binding energy. In
this context, a `polaron' describes an object that is not just a
charge coupled to a local lattice distortion, but which has a
complex spin and orbital structure, rather like the AFM/CO
correlations considered in Ref.~\cite{Yu08}.

An important result of Fig.~\ref{Figure3} is that there is neither a
discontinuity nor an abrupt change in the shape of spectra across
the insulator-to-metal transition. No hard gap opens on warming
through \TMI\ but the DOS is depleted in the vicinity of $\mu$ at
all measured temperatures. However, subtle and gradual changes in
the spectral shape are observed as a function of temperature. In the
insulating phase, on cooling towards the transition, ZBC decreases
as peaks/kinks develop at the flanks of the depletion, indicating a
spectral weight transfer from low to high energies. On further
cooling, in the metallic phase spectral weight increasingly builds
up at the chemical potential and the height of conductance peaks
slightly decreases, see Fig.~\ref{Figure3}(a). In terms of the
findings of Ref.~\cite{Yu08}, these results can be interpreted as
follows: In the insulating phase the AFM/CO correlation lifetime
increases upon cooling, but decreases as ferromagnetic order sets in
at $T<$\TMI. Although the short lifetime of the correlations
hindered us to directly image them in conductance maps, they are
manifest through the pseudogapped-like depletion of tunneling
conductance and the temperature evolution of the ZBC. It is
important to note that the calculations of Ref.~\cite{Yu08} were
performed in the clean limit. Quenched disorder is known to enhance
the colossal magnetoresistance and extends the region of the phase
diagram where it occurs~\cite{Dagotto08}, but our results call for a
systematic study of its effect on the AFM/CO correlations dynamics.

\begin{figure}[ttt!]
\includegraphics[width=0.5\textwidth]{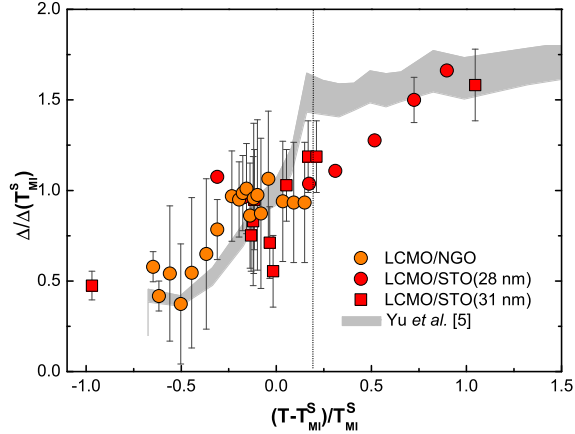}
\caption{\label{Figure4} Pseudogap energy (relative to its value at
$\mathrm{T_{MI}^{S}}$) vs. reduced temperature for LCMO/STO in red
(circles are data from this work, squares from Ref.~\cite{Seiro08}),
LCMO/NGO (orange) and results from the simulations in
Ref.~\cite{Yu08} (gray-shaded zone, covers the error). The
temperatures are given relative to $\mathrm{T_{MI}^{S}}$ for all
experimental data (see caption of Fig.~\ref{Figure2}), and relative
to the temperature of the local DOS minimum for the simulation.}
\end{figure}

The width of the depletion or pseudogap, $\Delta$, was obtained for
the curves in Fig.~\ref{Figure3} as the half-distance between the
two peaks (kinks) in the case of LCMO/STO(NGO) films. The pseudogap
values are summarized in Fig.~\ref{Figure4} for all
samples~\cite{Futinochi2}. The behavior of $\Delta$ as a function of
temperature is in qualitative agreement with the results of
Ref.~\cite{Yu08}, globally increasing on warming. In the vicinity of
\TMI, the pseudogap energy changes its slope but $\Delta$ further
increases in the high-temperature phase.

In conclusion, we have carried out a detailed analysis of the tunnel
spectra of manganite films that sheds light into the puzzling
relation between the local electronic properties and macroscopic
transport across the insulator-to-metal transition. Although both
the insulating and metallic phases present a pseudogapped normalized
conductance, spectral weight redistributes as a function of
temperature in such a way that the temperature evolution of the
macroscopic conductivity is tracked by the density of states at the
chemical potential. A similar behavior was predicted in
Ref.~\cite{Yu08}, where nanoscale spin/charge correlations were
found to be increasingly stable on cooling towards \TMI\ but their
lifetime is reduced with the onset of ferromagnetic order. In
addition, the predicted temperature evolution of the pseudogap
energy~\cite{Yu08} is consistent with the measured $\Delta(T)$. This
behavior is observed for highly and weakly strained films. Our
results strongly support that dynamical spin/charge correlations
play a preeminent role in the transport properties of colossal
magnetoresistive manganites.

The authors thank Y. Rong for useful discussions and the Swiss
National Science Foundation/MaNEP for financial support.


\begin{references}
\bibitem{barrier}
In our experiments the apparent workfunction ranged between 1 and
1.5\,eV.
\end{references}
\end{document}